\documentclass{nature}

\usepackage[dvips]{graphicx}

\def\gsim{\mathrel{\rlap{\lower 4pt \hbox{\hskip 1pt $\sim$}}\raise 1pt \hbox {$>$}}} \def\lsim{\mathrel{\rlap{\lower 4pt \hbox{\hskip 1pt $\sim$}}\raise 1pt \hbox {$<$}}}

\title{An asymmetric explosion as the origin of spectral \\ 
evolution diversity in type Ia Supernovae}

\author{
K. Maeda$^{1}$, S. Benetti$^{2}$, M. Stritzinger$^{3,4}$, 
F.~K. R\"opke$^{5}$, G. Folatelli$^{6}$,  J. Sollerman$^{7,4}$, \\ 
S. Taubenberger$^{5}$, K. Nomoto$^{1}$, 
G. Leloudas$^{4}$, M. Hamuy$^{6}$, 
M. Tanaka$^{1}$, P.~A. Mazzali$^{5,8}$, \\ 
N. Elias-Rosa$^{9}$ 
}

\begin{document}

\maketitle

{\footnotesize\it Published in Nature, 1 July 2010 issue.}

\begin{affiliations}
 \item Institute for the Physics and Mathematics of the Universe (IPMU), 
University of Tokyo, 5-1-5 Kashiwanoha, Kashiwa, Chiba 277-8583, Japan: 
keiichi.maeda@ipmu.jp.
 \item INAF - Osservatorio Astronomico di Padova, vicolo dell'Osservatorio 5, 
I-35122 Padova, Italy. 
 \item Carnegie Institute for Science, Las Campanas Observatory, 
Colina el Pino Casilla 601, La Serena, Chile. 
 \item Dark Cosmology Centre, Niels Bohr Institute, Copenhagen University, 
Juliane Maries Vej 30, 2100 Copenhagen \O, Denmark. 
 \item Max-Planck-Institut f\"ur Astrophysik, 
Karl-Schwarzschild-Stra{\ss}e 1, 85741 Garching, Germany. 
  \item Universidad de Chile, Departamento de Astronom\'ia, Casilla 36-D, Santiago, Chile. 
 \item The Oskar Klein Centre, Department of Astronomy, 
Stockholm University, AlbaNova, 10691 Stockholm, Sweden. 
 \item Scuola Normale Superiore, Piazza Cavalieri 7, 56127 Pisa, Italy. 
 \item Spitzer Science Center, California Institute of Technology, 1200 E. 
California Blvd., Pasadena, CA 91125, USA. 
\end{affiliations}

\clearpage 

\begin{abstract} 
Type Ia Supernovae (SNe Ia) form an observationally uniform class of 
stellar explosions, in that more luminous objects have smaller 
decline-rates$^{1}$. 
This one-parameter behavior allows 
SNe~Ia to be calibrated as cosmological `standard candles',
and led to the discovery of an accelerating Universe$^{2,3}$.
Recent investigations, however, have revealed that 
the true nature of SNe~Ia is more complicated. 
Theoretically, it has been suggested$^{4-8}$ that the initial thermonuclear 
sparks are ignited at an offset from the centre of the white-dwarf (WD) progenitor, 
possibly as a result of convection before the explosion$^{4}$. 
Observationally, the diversity seen in the spectral evolution 
of SNe~Ia beyond the  luminosity decline-rate relation
is an unresolved issue$^{9,10}$.
Here we report that the spectral diversity 
is a consequence of random directions from which an asymmetric explosion is viewed.
Our findings suggest that the spectral evolution diversity 
is no longer a concern in using SNe Ia as cosmological 
standard candles. Furthermore, this indicates that ignition at an
offset from the centre of is a generic feature of SNe~Ia. 
\end{abstract}

\clearpage

When a carbon-oxygen WD reaches  
a critical limit known as the Chandrasekhar mass 
($\sim 1.38~M_{\odot}$), 
its central density and temperature increase to a point where
a thermonuclear runaway is initiated. 
The thermonuclear sparks give birth to a subsonic deflagration flame, 
which at some point may make a transition to a supersonic detonation wave 
that leads to the complete disruption of the WD$^{11,12}$. 
The thermalization of  $\gamma$-rays produced from the 
decay of freshly synthesized radioactive $^{56}$Ni powers the  transient 
source, known as a SN~Ia$^{13,14}$. 
The relationship between the luminosity and the 
decline-rate parameter ($\Delta$m$_{15} (B)$, which 
is the difference between the $B$-band magnitude at peak
and that measured 15 days later) is interpreted to 
be linked to the amount of newly synthesized $^{56}$Ni (refs. 15, 16). 

SNe~Ia displaying a nearly identical photometric evolution can 
exhibit appreciably different 
expansion velocity gradients ($\dot v_{\rm Si}$) as inferred from the 
Si~II~$\lambda$6355 absorption feature$^{10,17}$.
More specifically, 
objects that show $\dot v_{\rm Si}$  $\gsim$ 70 km~s$^{-1}$ day$^{-1}$
are placed into the  high-velocity gradient  (HVG)  group, 
while those that show smaller gradients are 
placed in the low-velocity gradient (LVG) group. 
For normal SNe~Ia$^{18}$, which are the predominant 
population of the total SN~Ia sample 
and the main focus of this {\it Letter}, $\dot v_{\rm Si}$ is {\it not} 
correlated with $\Delta$m$_{15} (B)$ (ref. 10; Fig.~1a, 1b), 
thus raising a nagging concern regarding the `one-parameter' description. 

Late phase nebular spectra can be used to trace the distribution of the 
inner ejecta$^{19}$. 
Beginning roughly half a year after explosion, as the
ejecta expands, its density decreases  to the point where  
photons freely escape. 
Photons originating from the near/far side of the ejecta are
detected at a shorter/longer (blue-shifted/red-shifted) wavelength 
because of Doppler shifts.  
For SNe~Ia, emission lines related to [Fe II]~$\lambda$7155 
and [Ni II]~$\lambda$7378 
are particularly useful, as they are formed in the ashes of the deflagration flame$^{19}$. 
These lines show diversity in their central wavelengths -- 
blue-shifted in some SNe~Ia and red-shifted in others (see Fig.~1c) -- 
which provides evidence that the deflagration ashes, therefore the initial sparks, 
are on average located off-centre.  
The wavelength shift can be translated to a 
line-of-sight velocity ($v_{\rm neb}$) of the deflagration ashes. 

Figure 2 shows a comparison between $\dot v_{\rm Si}$ 
and $v_{\rm neb}$ for 20 SNe~Ia. 
Details regarding the data are provided in SI \S 1.
Although the diversities in these observables were discovered independently, 
Fig.~2 clearly shows that they are connected. 
Omitting the peculiar SN 2004dt 
(Fig. 2 Legend; SI \S1), all 6 HVG SNe 
show $v_{\rm neb} > 0$ km s$^{-1}$ (i.e., red-shifts), which means 
that these events are viewed from the direction opposite to the 
off-centre initial sparks.
The 11 LVGs display a wider distribution in
$v_{\rm neb}$ space, but are concentrated to negative 
values (i.e., blue-shifted), 
indicating that these events are preferentially viewed from the 
direction of the initial sparks. 
If HVG and LVG SNe were intrinsically 
distributed homogeneously as a function of $v_{\rm neb}$, 
the probability that just by chance (as a statistical fluctuation) all HVG SNe 
show $v_{\rm neb} > 0$ km s$^{-1}$ is merely 0.4\%; it is thus quite unlikely. 
Indeed, the probability that by chance 
6 HVG SNe are among the 7 SNe showing the largest red-shift 
in $v_{\rm neb}$ in our sample of 17 SNe is only 0.06\%. 

This finding strongly 
indicates that HVG and LVG SNe do not have intrinsic differences, 
but that this diversity arises solely from a viewing angle effect. 
Figure 3 shows a schematic picture. 
If viewed from the direction of the off-centre 
initial sparks, an SN~Ia appears as an LVG event at early phases 
and shows blue-shifts in the late-time emission-lines. 
If viewed from the opposite direction, it appears as an HVG event, and shows 
red-shifts at late phases. 

The number of HVG SNe is $\sim 35$\%$^{10}$ of the total number of HVG and LVG SNe. 
To explain this, the angle to the observer 
at which an SN changes its appearance from an LVG to an HVG 
is $\sim 105 - 110^{\circ}$, measured relative to the direction between the centre and 
the initial sparks. The velocity shift of $3,500$ km s$^{-1}$ in the distribution 
of the deflagration ashes, as derived for the normal SN~Ia 2003hv through a 
detailed spectral analysis$^{19}$, corresponds to 
$v_{\rm neb} \sim 1000$ km~s$^{-1}$ if viewed from this transition angle. 
The configuration then predicts that all LVG SNe should show 
$-3,500 < v_{\rm neb} < 1000$ km~s$^{-1}$, 
while all HVG SNe should be in the range 
$1000 < v_{\rm neb} < 3,500$ km s$^{-1}$ .
These ranges are
shown as arrows in Fig. 2, and  
provide a good match to the observations.

Figure 4a shows an example of a hydrodynamic model 
in which the thermonuclear sparks were ignited off-centre in a Chandrasekhar-mass WD$^{6}$ 
(an alternative way of introducing global asymmetries is 
double detonations in sub-Chandrasekhar-mass WDs$^{20}$).
Although this model has not been fine-tuned to 
reproduce the present finding, 
it does have the required generic features. 
The density distribution is shallow and extends to high velocity 
in the direction opposite to the initial sparks (Fig. 4b). 
Initially, the photosphere is at high velocity if viewed from this direction, 
as the region at the outer, highest velocities is still opaque. 
Later on, the photosphere recedes inwards faster in this opposite 
direction, owing to the shallower density gradient. 
As a result, the SN looks like an LVG if viewed from the offset direction, 
but like an HVG SN from the opposite direction (Fig. 4c), 
as in our proposed picture (Fig. 3). 

Our finding provides not only strong support for the asymmetric explosion 
as a generic feature,  
but also constraints on the still-debated deflagration-to-detonation transition. 
In this particular simulation, the change in appearance 
(as an HVG or an LVG SN) takes place rather abruptly around 
the viewing direction of $\sim 140^{\circ}$. 
Owing to the offset ignition, the deflagration flame propagates 
asymmetrically and forms an off-centre, shell-like region of 
high density deflagration ash.  
The detonation is ignited at an offset following the deflagration, but tries to expand 
almost isotropically.  
However, the angle between 
$0^{\circ}$ and $140^{\circ}$ is covered by the deflagration ash, 
into which the strong detonation wave (fueled by the unburned material 
near the centre of the WD) cannot penetrate. 
On the other hand, in the $140 - 180^{\circ}$ direction, the 
detonation can expand freely, creating a shallow density distribution. 
The `abrupt' change in appearance, as inferred by the observational data, 
is therefore a direct consequence of the offset models, controlled by 
the distribution of the deflagration ash. 
The `opening angle' of the transition is 
on the other hand dependent on the details of 
the explosion. 
To accurately model this according to our finding 
($\sim 105 - 110^{\circ}$ for the typical transition angle), 
either a smaller offset of the initial 
deflagration sparks or an earlier deflagration-to-detonation transition 
would be necessary. 
Such changes are also required to produce the typical 
velocity shift of $\sim 3,500$ km s$^{-1}$ in the distribution of the deflagration ash. 

Our proposed model unifies into a single scheme recent advances 
in both theoretical and observational studies of SNe~Ia - 
and it does not conflict with other results  
produced by spectral tomography$^{16,21}$ or
polarization measurements$^{22}$ (SI \S2). 
Our interpretation suggests that two SNe with very similar
light curve evolution may not necessarily produce 
exactly the same amount of $^{56}$Ni. 
They might in fact 
show somewhat different light curve evolution 
if viewed from the same direction (relative to the offset between the centre and the initial sparks), 
but the light curves look the same to an observer as 
$\Delta$m$_{15} (B)$ can change with viewing direction$^{5}$ (SI \S3). 
This situation is a natural consequence of observing a number of SNe Ia with 
a large variation in $^{56}$Ni production and in the viewing angles. 
Our finding regarding the explosion mechanism 
will lead to quantitative evaluation on the contribution of this random effect 
to the observed scatter in the SN~Ia luminosities beyond the one-parameter 
description$^{5}$, as compared to other systematic effects, such as 
the stellar environment$^{23}$.



\begin{addendum}
\item The authors thank Wolfgang Hillebrandt for discussions. 
This study is partly based on observations obtained at the Gemini Observatory, 
Chile (GS-2009B-Q-8, GS-2008B-Q-32/40/46), the Magellan Telescopes, 
Chile, and by ESO Telescopes at the La Silla or Paranal Observatories 
under programme 080.A-0516. 
This research made use of the {\it SUSPECT} archive, 
at the Department of Physics and Astronomy, University of Oklahoma. 
This work was supported by World Premier International Research Center
Initiative (WPI Initiative), MEXT, Japan. 
K. M. was supported by JSPS Grant-in-Aid for young scientists. 
S.B. acknowledges partial support from ASI contracts `COFIS'.
M.S. was supported by the National Science Foundation. 
F.K.R.\ was supported through the Emmy Noether Program of the German Research Foundation 
and by the Cluster of Excellence `Origin and Structure of the Universe'. 
G.F. and M.H. acknowledge support from Iniciativa Cientifica Milenio and CONICYT programs FONDECYT/FONDAP/BASAL.
J.S. is a Royal Swedish Academy of Sciences Research Fellow supported by 
the Knut and Alice Wallenberg Foundation. 
S.T. acknowledges support by the Transregional Collaborative Research Centre under the programme 
`The Dark Universe'. The Dark Cosmology Centre is funded by the Danish 
National Research Foundation. 
 \item[Author Contributions] S.B. and K.M. found the relation between the velocity 
gradient and the nebular velocity, initiated and organized the project. 
K.M. wrote the manuscript with the assistance of  M.S., J.S., G.F., and S.T.  
S.B. is responsible for the late-phase spectrum of SNe 1997bp. 
M.S., G.F., and M.H are responsible for acquisition and reduction of SNe 2007on, 2007sr, 
and 2009ab. 
F.K.R. and K.M. are responsible for the explosion simulation. 
All the authors contributed to discussions. 
 \item[Competing Interests] The authors declare that they have no
competing financial interests.
 \item[Correspondence] Correspondence and requests for materials
should be addressed to K.M. (email:\\ keiichi.maeda@ipmu.jp).
\end{addendum}

\clearpage
\begin{figure}
 \begin{center}
\includegraphics[width=0.45\textwidth]{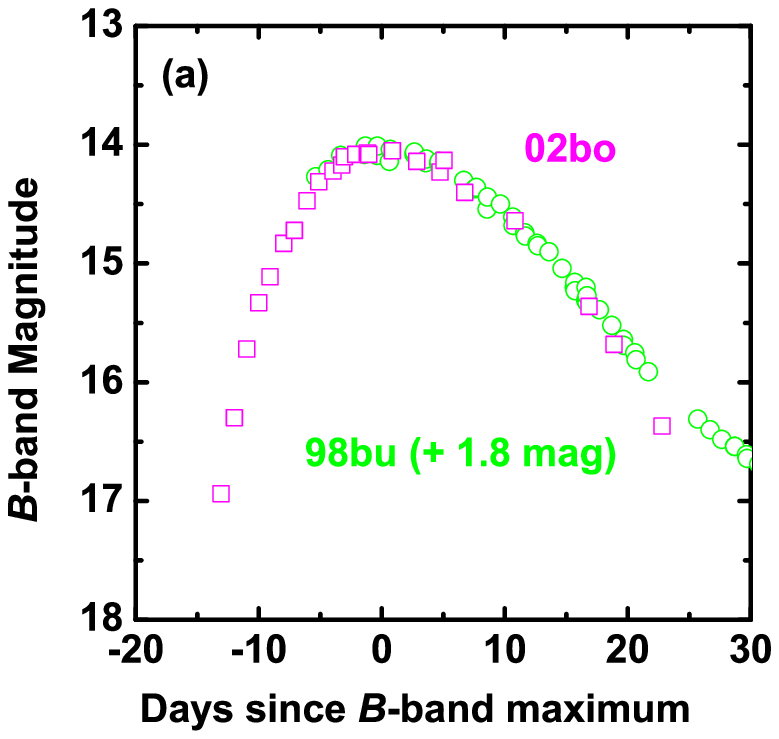}\\
\includegraphics[width=0.45\textwidth]{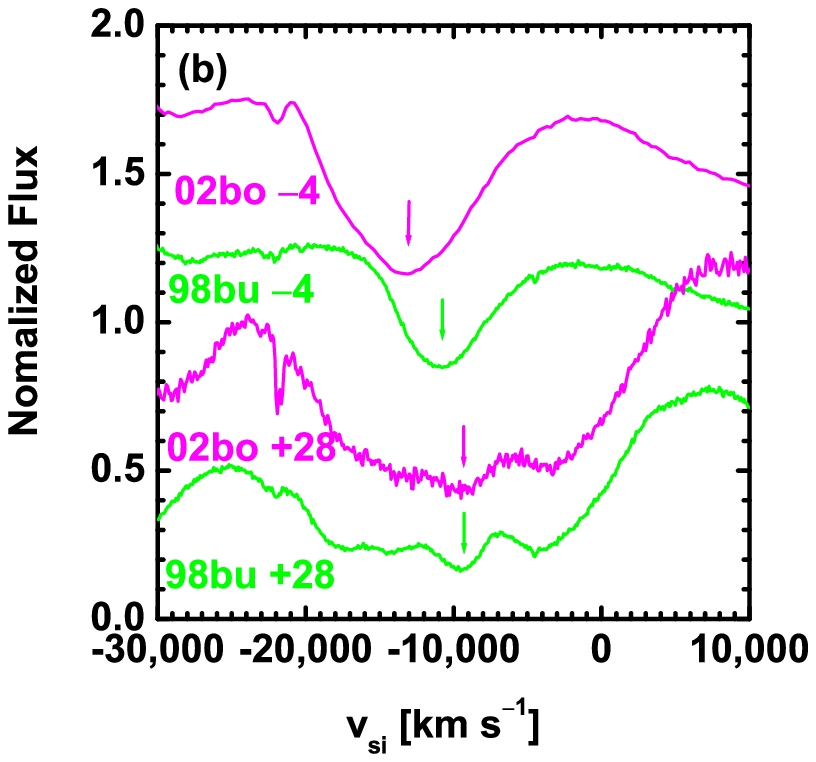}\\
\includegraphics[width=0.45\textwidth]{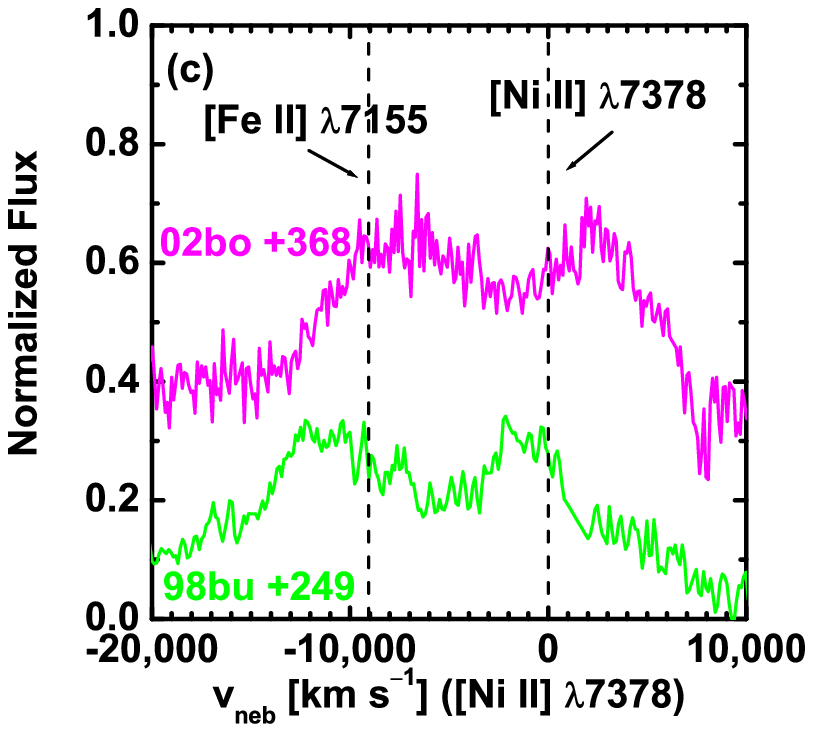}
 \end{center}
\end{figure}

\noindent
\sffamily\noindent\textbf {Figure 1}\hspace{2em}
{\sf {\bf Comparison between HVG SN Ia 2002bo and LVG SN Ia 1998bu. } 
The decline-rate parameter $\Delta$m$_{15} (B)$ is 
$1.16$ and $1.06$ mag for SNe 2002bo and 1998bu, respectively. 
{\bf a. } The $B$-band light curves$^{17,24}$. 
The magnitudes for SN 1998bu have been artificially 
shifted in the $y$-direction for presentation. 
{\bf b. } Si II~$\lambda$6355 at different epochs$^{17,24}$ 
(in days with respect to $B$-band maximum). 
SN 2002bo had initially a larger absorption velocity than 
SN 1998bu, but later its velocity approached that of SN 1998bu. 
The velocity evolved quicker and the velocity 
gradient ($\dot v_{\rm Si}$) is larger for SN 2002bo than for SN 1998bu. 
{\bf c. }  [Fe II]~$\lambda7155$ and [Ni II]~$\lambda7378$ 
in late-time spectra$^{21,25}$. 
The horizontal axis denotes the line velocity measured from the 
rest wavelength of [Ni II]~$\lambda$7378. The rest wavelengths of 
[Fe II]~$\lambda$7155 and [Ni II]~$\lambda$7378 are marked by dashed lines. 
These lines are red-shifted in SN 2002bo while blue-shifted in SN 1998bu. 
The wavelength shift indicates the line-of-sight velocity of 
the deflagration ashes ($v_{\rm neb}$) ($v_{\rm neb} < 0$ km s$^{-1}$, 
i.e., the blue-shift, if the material is moving toward us). 
These are the strongest lines among those emitted from the deflagration 
ash according to the previous analysis of late-time emission lines$^{19}$. 
Indeed, there are stronger lines which do not show Doppler shifts,
e.g., [Fe III]~$\lambda$4701; they however 
do not trace the distribution of the deflagration ash$^{19}$ 
(see also SI \S 1) and are thus not useful in the present study. 
}

\clearpage
\begin{figure}
 \begin{center}
  \includegraphics[width=0.8\textwidth]{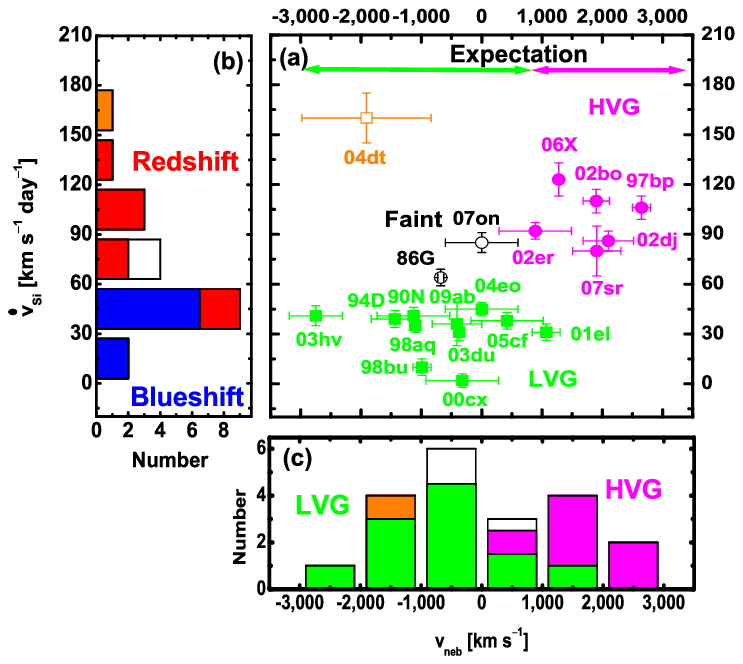}
 \end{center}
\end{figure}

\sffamily\noindent\textbf {Figure 2}\hspace{2em}
{\sf {\bf Relations between the features in early- and late-phases. } 
{\bf a.} 
Early-phase velocity gradient ($\dot v_{\rm Si}$: vertical axis) 
as compared to late-phase emission-line shift velocity ($v_{\rm neb}$: horizontal axis) 
for 20 SNe Ia. The errors are for $1\sigma$ in fitting the velocity evolution for 
$\dot v_{\rm Si}$, 
while for
$v_{\rm neb}$ the errors are from differences 
in measurement between different emission lines (see Sup \S1). 
LVG SNe and HVG SNe are shown 
by green squares and magenta circles, respectively. 
SNe 1986G and 2007on are classified as `faint and fast-declining' 
SNe Ia$^{26-28}$ (black open circles). 
SN 2004dt (orange square) 
is an HVG SN according to the value of $\dot v_{\rm Si}$, 
but displayed peculiarities in the 
late-time spectrum$^{29}$ (SI \S1) and in polarization 
measurements$^{22}$ (SI \S2), and the value for 
$\dot v_{\rm Si}$ is exceptionally large as compared to other HVG SNe. 
These suggest that SN 2004dt is an outlier and the origin of its large 
$\dot v_{\rm Si}$ is probably different from that of other SNe Ia. 
The two arrows on top indicate the regions where HVG and LVG SNe are expected, 
based on a simple kinematic interpretation (see main text). 
{\bf b. } Number distribution of 20 SNe as a function of $\dot v_{\rm Si}$. 
White and orange areas are for faint SNe and SN 2004dt. The remaining 
SNe are marked depending on whether they show a blue-shift ($v_{\rm neb} < 0$ 
km s$^{-1}$: blue area) or red-shift ($v_{\rm neb} > 0$ km s$^{-1}$: red area) 
in their late-time spectra. 
{\bf c. } 
Number distribution of 20 SNe as a function of $v_{\rm neb}$. 
}

\clearpage
\begin{figure}
 \begin{center}
  \includegraphics[width=0.6\textwidth]{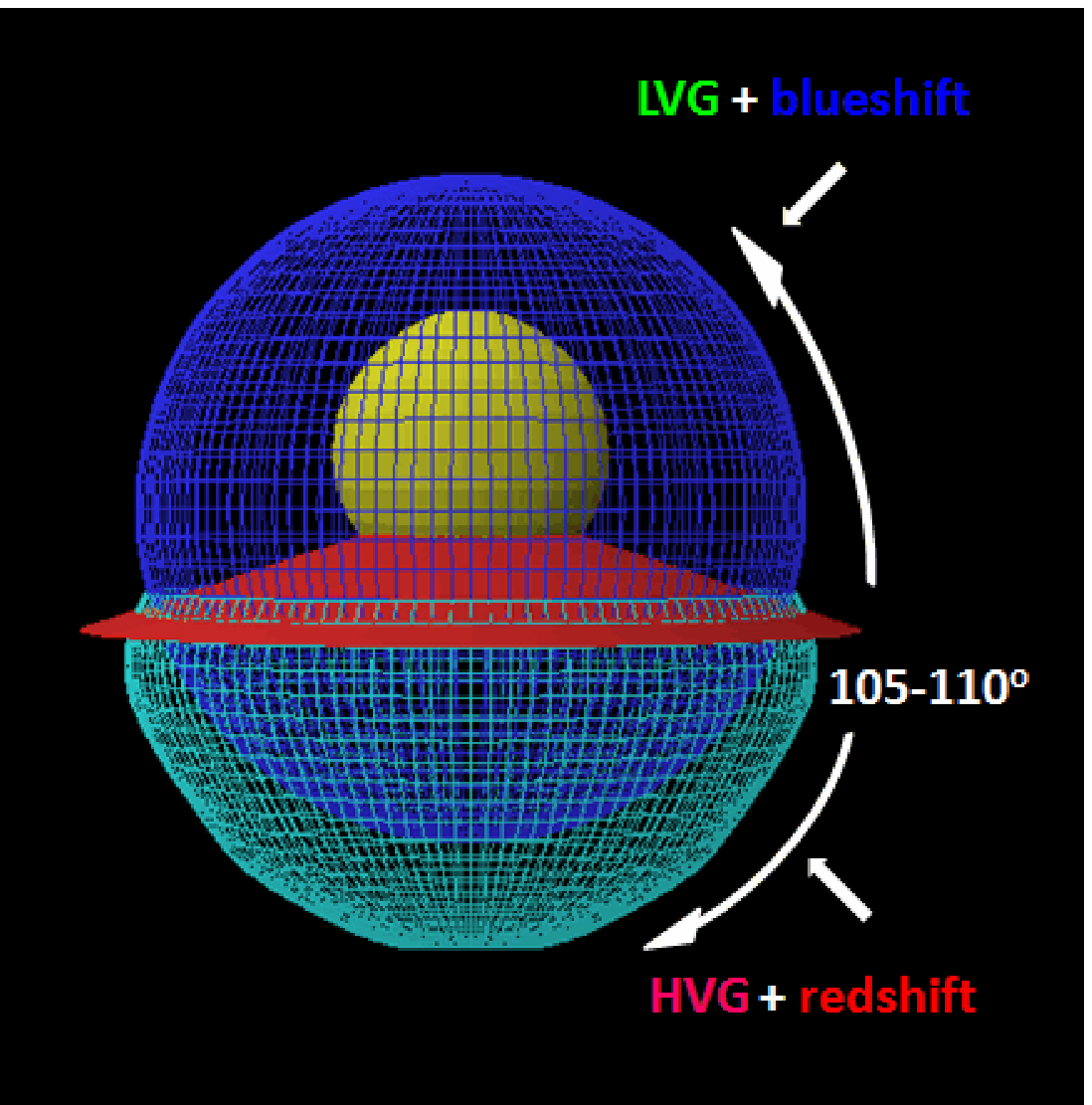}
 \end{center}
\end{figure}
\sffamily\noindent\textbf {Figure 3}\hspace{2em}
{\sf {\bf A schematic picture of the structure of SN Ia ejecta. } 
This configuration simultaneously explains 
the relative fractions 
of HVG and LVG SNe Ia, and 
the relation between $\dot v_{\rm Si}$ and $v_{\rm neb}$. 
It is also consistent with the diversity in $v_{\rm neb}$ (ref. 19). 
The ashes of the initial deflagration sparks 
are shifted with respect to the center of the SN ejecta by 
$\sim 3,500$ km s$^{-1}$ (yellow: 
Although expressed by a spherical region for presentation, it may well 
have an amorphous shape owing to the hydrodynamic instability in the deflagration flame$^{7}$). 
This region is rich in stable $^{58}$Ni with a small amount of radioactive $^{56}$Ni, and 
emits [Fe II]~$\lambda$7155 
and [Ni II]~$\lambda$7378 at late phases$^{19}$ (SI \S 1). 
The outer region is the later detonation ash 
responsible for the early-phase Si II~$\lambda$6355 absorption. 
This region is roughly spherically distributed (blue), but extends to 
the outer region in the direction opposite to the deflagration 
ashes (cyan). Although the detonation can produce $^{56}$Ni which decays 
into $^{56}$Fe, it has been argued that 
these regions (blue and cyan) are not main contributors to 
[Fe II]~$\lambda$7155 and [Ni II]~$\lambda$7378 at late phases$^{19}$ (SI \S 1). 
A putative observer would view this explosion as 
an LVG or HVG SN, depending on the direction of the observer, as divided by 
a specific angle (red), which is $\sim 105 - 110^{\circ}$. 
This angle is derived from the 
relative numbers of HVG and LVG SNe (see main text); 
the fraction of HVG SNe to the total number of 
HVG and LVG SNe $\sim 35$\% (ref. 10), 
as was also supported by a larger sample$^{30}$ 
with more than 100 SNe (using the observed trend that the HVG SNe show 
higher velocities than LVG objects at early times). 
}

\clearpage
\begin{figure}
 \begin{center}       
  \includegraphics[width=0.45\textwidth]{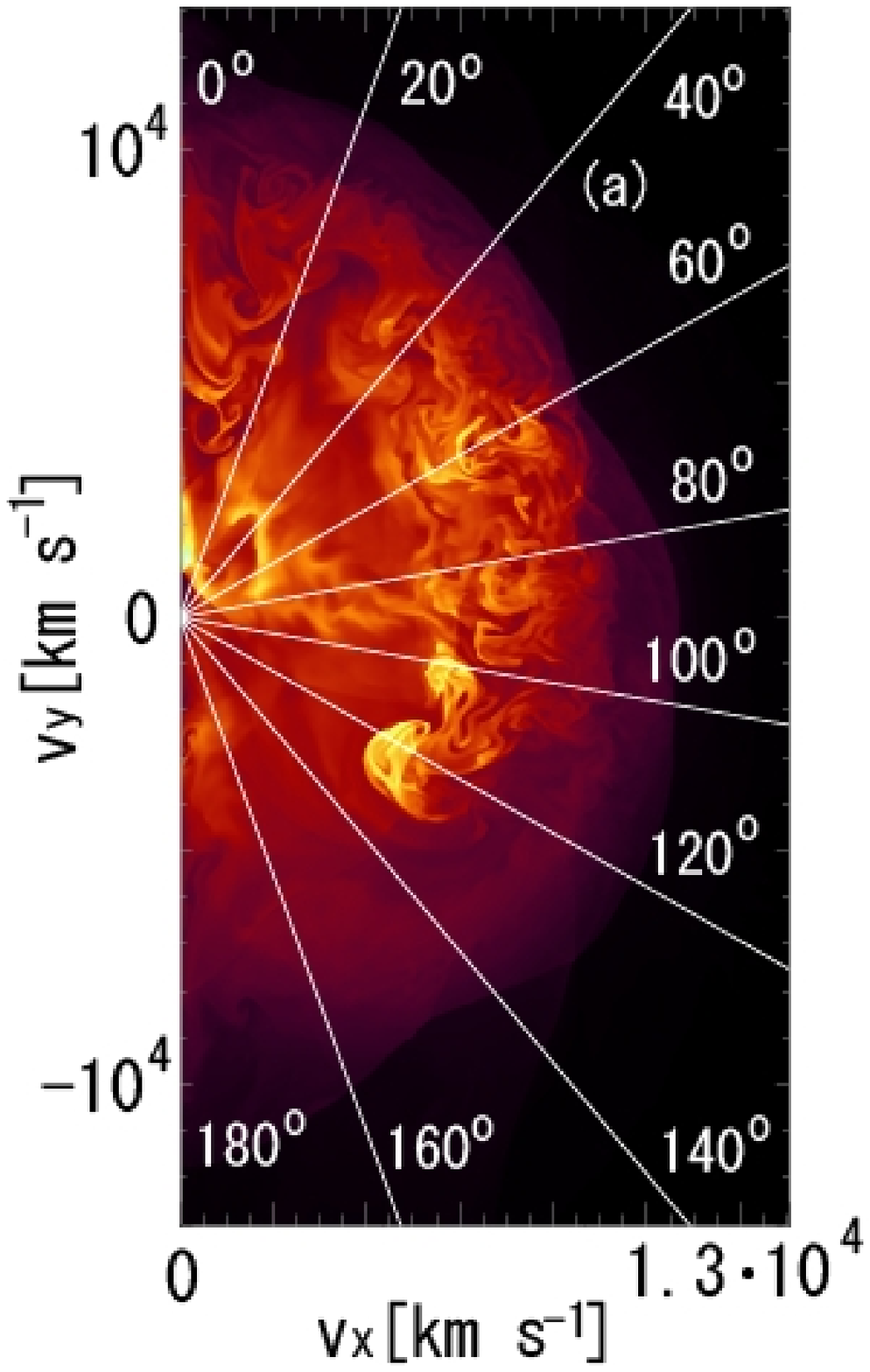}\\
  \includegraphics[width=0.45\textwidth]{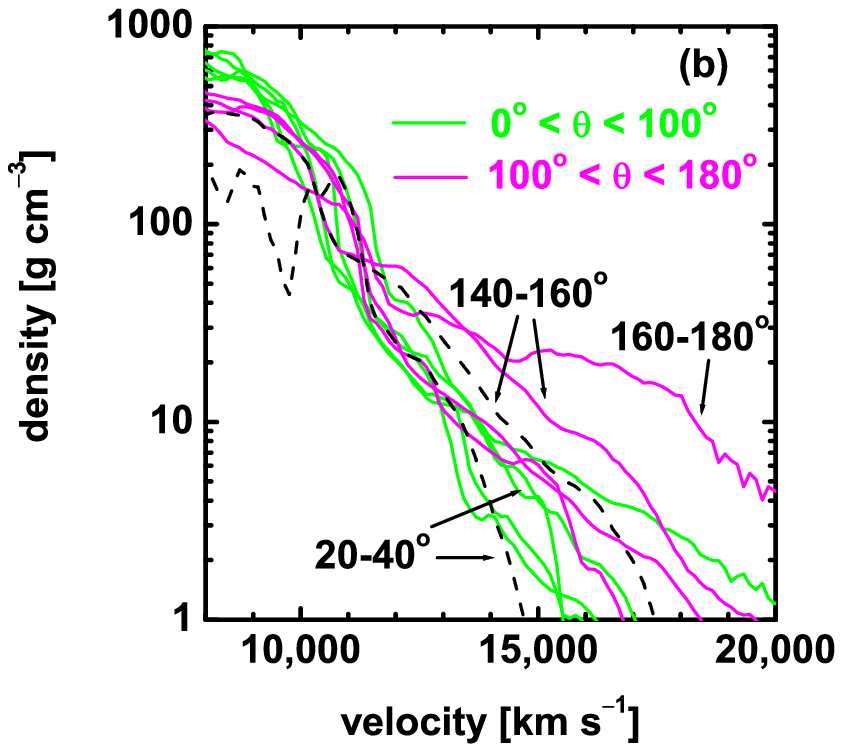}
  \includegraphics[width=0.45\textwidth]{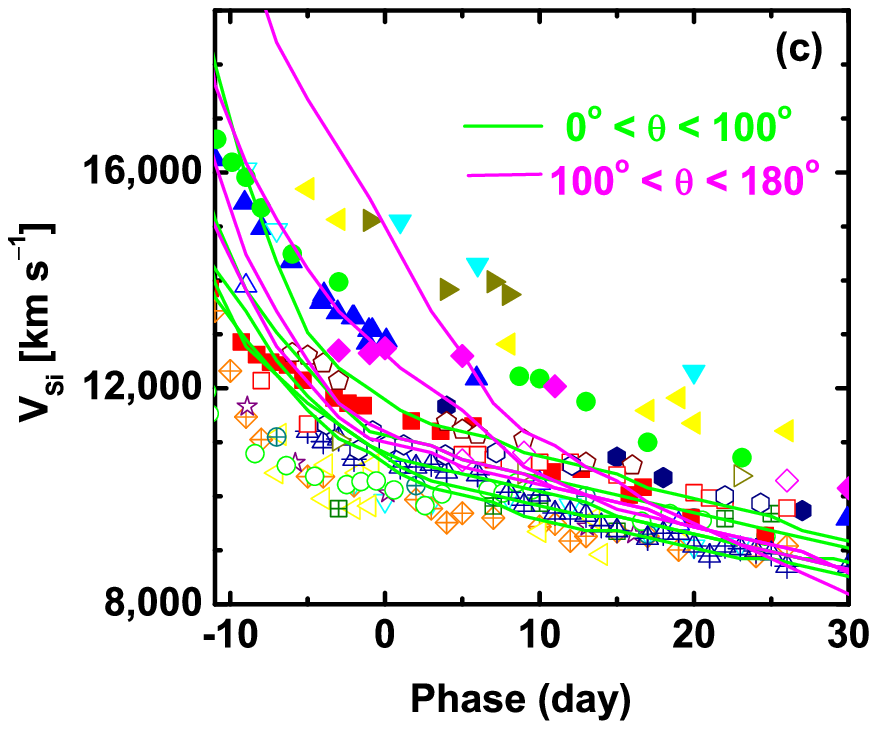}
 \end{center}
\end{figure}
\sffamily\noindent\textbf {Figure 4}\hspace{2em}
{\sf {\bf Expectations from a hydrodynamic explosion model. } 
{\bf a. } 
Cross section of the density distribution of an offset ignition 
model$^{6}$ (similar to a model in ref. 5). It is shown at 10 seconds 
after the ignition, when the homologous expansion is already reached. 
In this model, the deflagration sparks were ignited 
offset, within an opening angle of 45$^{\circ}$ and 
between $0 - 180$ km from the centre of a WD whose radius is $\sim 2,000$ km. 
The deflagration products are distributed in the high-density offset shell, 
which covers 
$0 - 140^{\circ}$ in this particular model. 
This resulting velocity shift is $\sim 8,000$ km s$^{-1}$ in this model, 
which is larger than the observational requirement 
($\sim 3,500$ km s$^{-1}$). 
{\bf b. } Radial density distribution of the same model 
for several directions (where the angle is measured from the direction of the offset sparks). 
The magenta lines show the density distribution for the angle in $100 - 180^{\circ}$, 
roughly corresponding to the putative directions for which a SN looks like 
an HVG SN inferred from the observational data (Fig. 3). 
Also shown is the density distribution multiplied by the mass fraction of Intermediate 
mass elements, which gives a rough distribution of Si, for two directions ($20 - 40^{\circ}$ 
and $140 - 160^{\circ}$). It is seen that the distribution of Si roughly follows the density 
distribution. 
{\bf c. } Model photospheric velocity evolution 
for several directions (lines) as compared 
to the observed Si II~$\lambda$6355 absorption line evolution$^{10}$ 
(filled and open symbols for HVG and LVG SNe, respectively; 
the phase in days with respect to $B$-band maximum). 
The position of the photosphere is estimated by integrating 
the optical depth with a constant opacity along each direction.  
Using the Si distribution for the velocity estimate provides a similar result. 
}

\clearpage
{\large Supplementary Information}

\section{Supernova Sample and Notes for Individual SNe}
Supplementary Table 1 summarizes the data of the SNe used in the present study. 
The values of the velocity gradients ($\dot v_{\rm Si}$) are obtained from the 
literature when available. 
They are mostly drawn from a previous compilation$^{10}$ 
(see Supplementary Tab. 1 for 
the sources of the observations), with values for additional SNe 
obtained from the references listed in Supplementary Table 1.  
For SN~1998aq, we have measured $\dot v_{\rm Si}$ from 
the published spectra$^{38}$. 

The line-of-sight velocity of the deflagration ash is derived as follows. 
Late-time spectra of SNe Ia show various forbidden lines from Fe-peak 
elements. These lines can be divided into two groups.
The first group requires intense 
heating from radioactive $^{56}$Ni $\to$ $^{56}$Co $\to$ $^{56}$Fe decays 
and low material density (e.g., [Fe III]~$\lambda$4701), 
while the second group requires a low heating rate and 
high material density (e.g., [Fe II]~$\lambda$7155 and [Ni II]~$\lambda$7378). 
Considering general properties of the emission process$^{19}$, 
the former lines are argued to be preferentially emitted from the detonation ash 
(because it is at relatively low density with a large amount of $^{56}$Ni), 
while the latter lines are formed in the deflagration ash 
(because it is at high density with a small amount of $^{56}$Ni). 
These two groups of lines show mutually different properties in observations, 
strengthening the interpretation that they are emitted from different regions: 
The `deflagration ash'-lines show the diversity in Doppler shift (not only 
in [Fe II]~$\lambda$7155 and [Ni~II]~$\lambda$7378$^{61, 62}$), 
while the `detonation ash'-lines show virtually no Doppler shifts. 
This property led to the conclusion that the deflagration ash is 
on average located offset, while the detonation ash is distributed roughly spherically$^{19}$. 

For $v_{\rm neb}$ (the line-of-sight velocity of the deflagration ash), 
we have therefore measured the wavelength shift in 
[Ni II]~$\lambda$7378 and in [Fe II]~$\lambda$7155, and defined 
$v_{\rm neb}$ as the mean value of them 
(see Supplementary Tab. 1 for the source of the observational data). 
The error bars are taken to be 
the difference in the measurements for the two lines. 
In all SNe except for SN 2004dt, the difference in these two measurements 
is at most $600$ km s$^{-1}$. For SNe Ia in which either of 
[Fe II]~$\lambda$7155 or [Ni II]~$\lambda$7378 
is weak and unidentified, we measure $v_{\rm neb}$ only from the other single line, 
with a conservative error of $600$ km s$^{-1}$. 
For SNe~2007on, 2007sr, and 2009ab, 
we have measured $\dot v_{\rm Si}$ and $v_{\rm neb}$ 
from our own spectra taken at the 
Gemini South, the Magellan, and the ESO (La Silla, Paranal) telescopes. 
The late-time spectrum 
of SNe~1997bp has also been obtained by us and used 
to measure $v_{\rm neb}$. 
The 20 SNe in Supplementary Table 1 are all the objects we have found 
for which both $\dot v_{\rm Si}$ and $v_{\rm neb}$ are reliably available to date. 
Also shown in the table are the light-curve decline-rate parameters
$\Delta m_{15} (B)$. 

Supplementary Table 1 also lists the epoch (after $B$-band maximum) 
of the late-time spectra from which 
$v_{\rm neb}$ is measured. 
For most of the SNe Ia, it is at least 200 days after 
$B$-band maximum, and thus errors caused by the blending of 
additional lines to this feature (e.g., permitted lines emitted at relatively early phases) 
in measuring $v_{\rm neb}$ can be avoided. 
The features at $\sim 7,000 - 7,500$~\AA, which we interpret to be dominated by 
[Fe II]~$\lambda$7155 and [Ni II]~$\lambda$7378, do not show significant evolution 
at $\gsim 100 - 150$ days after $B$-band maximum$^{19}$. 

It has been shown that normal SNe~Ia$^{18}$ (which constitute $\gsim 80$\% 
of the whole SN~Ia population), 
faint 1991-bg like SNe~Ia$^{26}$, and bright 1991T-like SNe~Ia$^{63,64}$ 
show different properties in $\dot v_{\rm Si}$$^{10}$. 
Normal SNe~Ia show a diversity in $\dot v_{\rm Si}$, which is apparently 
{\em not} 
correlated with $\Delta m_{15} (B)$. Faint 1991bg-like SNe~Ia, which are 
characterized by a rapid fading (i.e., large $\Delta m_{15} (B)$), always 
show large $\dot v_{\rm Si}$. Bright 1991T-like SNe~Ia 
with small $\Delta m_{15} (B)$ always show small $\dot v_{\rm Si}$. 
In deriving $v_{\rm neb}$, we have noticed that the identification of 
[Fe II]~$\lambda$7155 and [Ni II]~$\lambda$7378 is robust for normal SNe Ia, 
but tends to be a subject of possible misidentification for the other subclasses 
(then such SNe Ia are omitted from our analysis): 
Faint 1991bg-like SNe Ia tend to show probable [Ca II]~$\lambda$7291, 7324
$^{27}$. Bright 1991T-like SNe Ia tend to show a broad single peak, 
possibly indicating that [Fe II]~$\lambda$7155 and 
[Ni II]~$\lambda$7378 are broad and mutually blended. 
These indicate that the ejecta structure (density, temperature, and composition) 
of faint and bright SNe Ia is intrinsically different from normal ones. 
This is a reason why our sample is mainly composed of normal SNe Ia. 
Luckily, omitting a large fraction of faint/bright SNe Ia is not important 
for our present analysis, since 
we are interested in the spectral diversity beyond the one-parameter 
$\Delta m_{15} (B)$ description, which is a problem only in normal SNe Ia. 

For normal SNe~Ia, we categorize HVG SNe and LVG SNe according to
$\dot v_{\rm Si}$,  with the division line at $70$ km s$^{-1}$ day$^{-1}$. 
According to its $\dot v_{\rm Si}$, SN~2004dt is an HVG. 
However, its peculiar observational features 
suggest that it is an outlier, and the origins of its high velocity gradient 
and the negative $v_{\rm neb}$ are  
likely different than the one for other HVGs (see Fig.~2 caption). 
One of the peculiar features of SN~2004dt 
appears in its late-time spectra. 
Supplementary Fig.~1 shows a late-time spectrum of 
SN~2004dt compared to the prototypical faint SN~1991bg and to the 
normal HVG SN~2007sr. 
The spectrum of SN~2004dt provides a good match to that of the faint SN~1991bg. 
Both the intensity and the width of the lines at $\sim 7,000 -
7,500$~\AA\ are similar for these two SNe.
In normal HVG SN~2007sr, the features at $7,000 - 7,500$~\AA\ 
are much fainter than for SNe~2004dt and 1991bg. 
This indicates that in the case of SN~2004dt the feature is probably
contaminated by other lines, most likely [Ca~II]~$\lambda\lambda$7291, 7324 
(note that SN 2004dt shows a larger error in $v_{\rm neb}$ than the other SNe Ia). 
In addition, the strongest lines at late phases, i.e., 
the [Fe~III] blend at $\sim 4,700$~\AA\ 
and the [Fe~II] and [Fe~III] blend at $\sim 5,250$~\AA\ 
have similar ratios in SN~2004dt and  
1991bg; these ratios have been noticed to be peculiar$^{29}$. 
The normal HVG SN~2007sr has broader lines, and different line ratios. 
These similarities between the `normal' SN~2004dt and the faint 
SN~1991bg at late phases, which have not been noticed previously, 
suggest that these explosions might be closely 
related to one another. 
SN~2004dt may represent a new class of SNe~Ia$^{65}$ 
which have SN~1991bg-like  
features in the late-time spectrum, but are more energetic and brighter.

\begin{table*}
{\footnotesize
\begin{center}
Supplementary Table 1: Supernovae Sample{}
\begin{tabular}{ccccccc}
\hline
\hline
SN & $\Delta m_{15} (B)$ & $\dot v_{\rm Si}$ & $v_{\rm neb}$ 
& Epoch & Class$^{8,24}$ & References\\
&  & (km s$^{-1}$ day$^{-1}$) & (km s$^{-1}$) & (day) & \\
\hline
1986G & $1.81 \pm 0.07$ & $64 \pm 5$ & $-680 \pm 50$ & 257 & HVG/Faint & 1, 31, 32\\
1990N & $1.07 \pm 0.05$ & $41 \pm 5$ & $-1130 \pm 600$ & 280 & LVG & 1, 10, 33, 34\\
1994D & $1.32 \pm 0.05$ & $39 \pm 5$ & $-1440 \pm 390$ & 306 & LVG & 1, 34, 35\\
1997bp & $0.97 \pm 0.2$ & $106 \pm 7$ & $2650 \pm 150$ & 300 & HVG & 10, 36\\
1998aq & $1.12 \pm 0.05$ & $35 \pm 4$ & $-1100 \pm 50$ & 241 & LVG & 37, 38\\
1998bu & $1.06 \pm 0.05$ & $10 \pm 5$ & $-990 \pm 150$ & 329 & LVG & 1, 10, 24, 25, 39\\
2000cx & $0.93 \pm 0.04$ & $2 \pm 4$ & $-320 \pm 600$ & 147 & LVG/peculiar & 40-43\\
2001el & $1.13 \pm 0.04$ & $31 \pm 5$ & $1070 \pm 230$ & 398 & LVG & 44-46\\
2002bo & $1.16 \pm 0.06$ & $110 \pm 7$ & $1900 \pm 220$ & 368 & HVG & 17, 21\\
2002dj & $1.08 \pm 0.05$ & $86 \pm 6$ & $2100 \pm 420$ & 275 & HVG & 47\\
2002er & $1.32 \pm 0.03$ & $92 \pm 5$ & $890 \pm 600$ & 216 & HVG & 48, 49\\
2003du & $1.02 \pm 0.05$ & $31 \pm 5$ & $-370 \pm 50$ & 377 & LVG & 50, 51\\
2003hv & $1.61 \pm 0.02$ & $41 \pm 6$ & $-2750 \pm 440$ & 320 & LVG & 52\\
2004dt & $1.21 \pm 0.05$ & $160 \pm 15$ & $-1910 \pm 1070$ & 152 & HVG/peculiar & 29\\
2004eo & $1.45 \pm 0.04$ & $45 \pm 4$ & $0 \pm 600$ & 228 & LVG & 53, 54\\
2005cf & $1.12 \pm 0.03$ & $35 \pm 5$ & $420 \pm 600$ & 267 & LVG & 55-58\\
2006X & $1.31 \pm 0.05$ & $123 \pm 10$ & $1280 \pm 20$ & 277 & HVG & 59\\
2007on & $1.62 \pm 0.01$ & $85 \pm 6$ & $0 \pm 600$ & 356 & HVG/Faint & This work \\
2007sr & $0.93 \pm 0.02$ & $80 \pm 15$ & $1910 \pm 400$ & 256 & HVG & 60 \\
2009ab & $1.20 \pm 0.02$ & $36 \pm 13$ & $-410 \pm 410$ & 278 & LVG & This work \\
\hline         
\end{tabular}
\end{center}
}
\end{table*}

\clearpage

\section{Other Observational Constraints}
In abundance `tomography'$^{16,21,66-68}$, 
a temporal sequence of spectra of individual SNe~Ia 
are used to infer the distribution of different elements through the SN ejecta, 
assuming the density structure of a spherically symmetric 
explosion model$^{13}$. 
From this type of analysis, it has been indicated that 
the abundance distribution is generally a function of 
$\Delta m_{15} (B)$. 
The difference between the HVG and LVG SNe, not related 
to $\Delta m_{15} (B)$, is mainly on the extent of the 
Si-rich layer$^{16}$ and on the photospheric velocity$^{66}$, 
which are explained by our proposed scenario (main text).

On the other hand, the spatial extent of the $^{56}$Ni-rich region does 
not seem to be dependent on whether it is a HVG or a LVG SN$^{16}$. 
This could provide a constraint on the ejecta asymmetry. 
In the offset explosion model, the spatial extent of the $^{56}$Ni region, 
as well as 
the density structure at the outer edge of that region, are not sensitive to 
the direction, despite the initially large asymmetry in the ignition$^{6}$. 
This stems from the nature of the propagation of the detonation wave as 
described in the main text; unlike the deflagration flame, 
the detonation tries to expand isotropically, 
producing roughly spherically distributed $^{56}$Ni.  
This region is not sensitively affected 
by the existence of the deflagration ashes, which is essential in determining 
the structure of Si-rich region. 
As a result, the spatial extent of the $^{56}$Ni-rich region is 
mainly controlled by the different amounts of $^{56}$Ni produced 
in the explosion, 
and the viewing angle dependence could add some diversity at most as 
a secondary effect$^{6}$; 
this is consistent with the observational indications$^{16}$. 
 
The asymmetric distribution of the outermost layer may 
imprint its signature in polarization measurements, which may be linked 
to the velocity gradient$^{69}$.
The polarization of the Si~II line is correlated with 
$\Delta m_{15} (B)$$^{22}$, but only for LVG SNe (Supplementary Fig.~2). 
HVG SNe generally show larger polarization than LVG SNe$^{70,71}$ but
they  clearly do not follow this trend.
A global one-sided asymmetry as in the present interpretation would produce 
relatively low continuum polarization and relatively high line polarization at 
Si~II~$\lambda$6355$^{72}$. 
In our proposed scenario the global asymmetry is of smaller degree
than in an extremely asymmetric 
model producing Si~II polarization of $\sim 1$\%$^{72}$, 
thus this is likely 
not a major contributor to the observed polarization. 
Alternatively, it has been suggested, 
based on the correlation between the Si~II polarization 
and $\Delta m_{15} (B)$, 
that the observed Si~II polarization could be a measure of the 
thickness of the outer layer above the $^{56}$Ni-rich region, in which the 
local inhomogeneity, e.g., a few relatively dense blobs, 
is assumed to be a source of polarization. 
In this interpretation, the large Si~II polarization in HVG SNe 
could be a consequence of an extended outer layer 
in the direction opposite to the initial sparks. 

\clearpage
\section{Viewing-Angle Effect on the Light Curve}

It has been suggested that if the ejecta are asymmetric, 
$\Delta m_{15} (B)$ is dependent on the direction to the observer$^{5}$. 
Supplementary Fig.~3 shows the comparison between $\Delta m_{15} (B)$ 
and $v_{\rm neb}$ for SNe~Ia. 

In the same figure, 
the possible effect of the viewing angle on the light curve shape, and 
the implication on the scatter in the luminosity calibration 
beyond the one-parameter description, are schematically 
illustrated. For example, the LVG SN~1998bu and the HVG SN~2002bo had similar 
$\Delta m_{15} (B)$ values, and this could be interpreted as follows. 

These two SNe would indeed have intrinsically different amounts of 
$M$($^{56}$Ni) and accordingly different luminosity. 
Because of different $M$($^{56}$Ni), 
$\Delta m_{15} (B)$ would be different if viewed from the same direction. 
Assuming that (1) $\Delta m_{15} (B)$ would be smaller and the luminosity is 
thus larger for LVG SN~1998bu than for HVG SN~2002bo, for a putative 
observer at the same direction, and 
that (2) $\Delta m_{15} (B)$ appears 
larger for an observer closer to the offset direction, as 
a viewing angle effect. 
Then, these two SNe would show similar $\Delta m_{15} (B)$ as observed, 
despite the intrinsic difference in luminosity, 
since the viewing direction is closer to the offset direction for SN~1998bu. 

This effect would produce a scatter in the luminosity of SNe~Ia, around the 
standard luminosity determined by $M$($^{56}$Ni). 
Pairs of SNe with similar $\Delta m_{15} (B)$, as a result of 
the viewing angle originating from SNe with different amounts of $M$($^{56}$Ni), 
are naturally expected if we observe a large number of SNe~Ia (Sup. Fig.~3) 
for the following two reasons: 
(1) SNe~Ia do indeed show large variations in $M$($^{56}$Ni), and 
(2) the viewing angles should display large variations. 

A question for this interpretation is whether any indication of such 
an effect is seen in the data. Supplementary Fig.~3 shows that 
there is no clear (but perhaps a marginal) correlation between 
$\Delta m_{15} (B)$ and $v_{\rm neb}$. 
According to the prediction$^{5}$ of the 
viewing-angle effect on $\Delta m_{15} (B)$ 
for models similar to the one shown in 
the main text, the observed $\Delta m_{15} (B)$ could vary 
by $\sim 0.4$~mag depending on the direction to the observer (schematically 
shown in Sup. Fig.~3). 
This is smaller than the intrinsic variation in $\Delta m_{15} (B)$ 
for different $M$($^{56}$Ni), 
and thus such an effect is difficult to notice in Supplementary 
Fig.~3, as is consistent with the low correlation in the present data. 
Any marginal correlation between $\Delta m_{15} (B)$ and $v_{\rm neb}$ 
may already hint that such an effect is indeed there, 
but a larger number of SNe~Ia 
is necessary to test this possibility with statistical significance.

\clearpage
\section{Supplementary Figure 1}

 \begin{center}
  \includegraphics[width=0.8\textwidth]{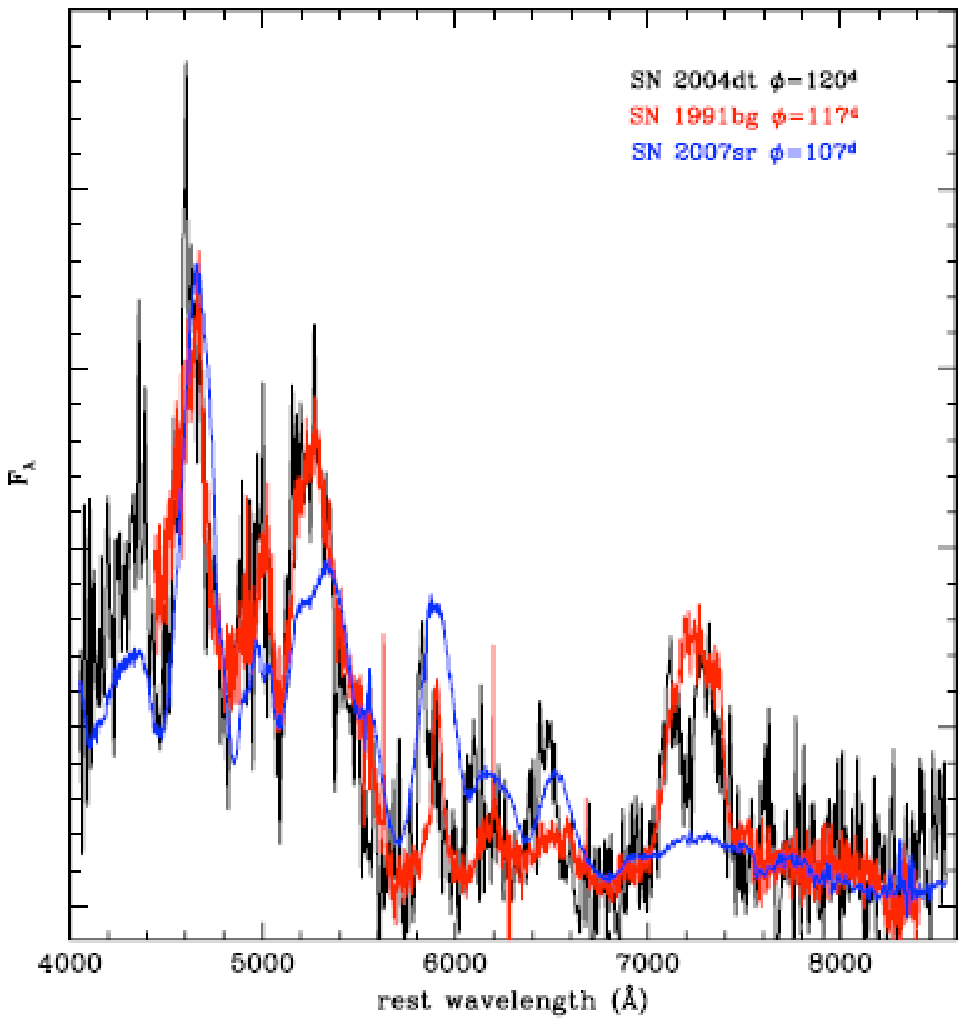}
 \end{center}

Spectrum of SN~2004dt$^{29}$ at 120 days after $B$-band maximum, 
compared to those of the faint SN~1991bg$^{27}$ and of the 
normal HVG SN~2007sr at similar epochs. 

\clearpage
\section{Supplementary Figure 2}

 \begin{center}
  \includegraphics[width=0.8\textwidth]{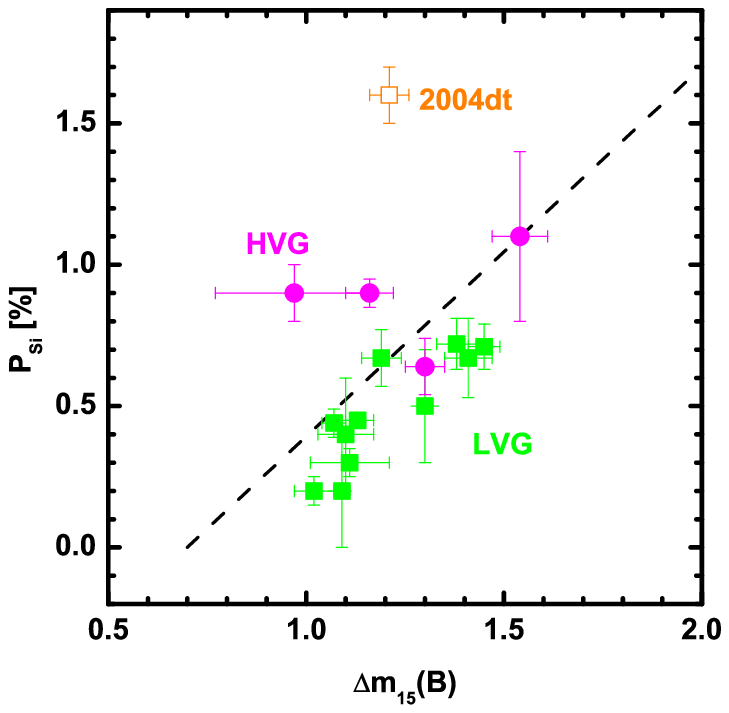}
 \end{center}

Si~II~$\lambda$6355 line polarization for SNe~Ia, 
as a function of the decline-rate parameter$^{22,70,71}$. 
The dashed line shows a linear fit to the data excluding 
SN~2004dt$^{22}$. 
The colors of the symbols are the same as in Fig.~2 
of the main text. 

\clearpage
\section{Supplementary Figure 3}

 \begin{center}
  \includegraphics[width=0.8\textwidth]{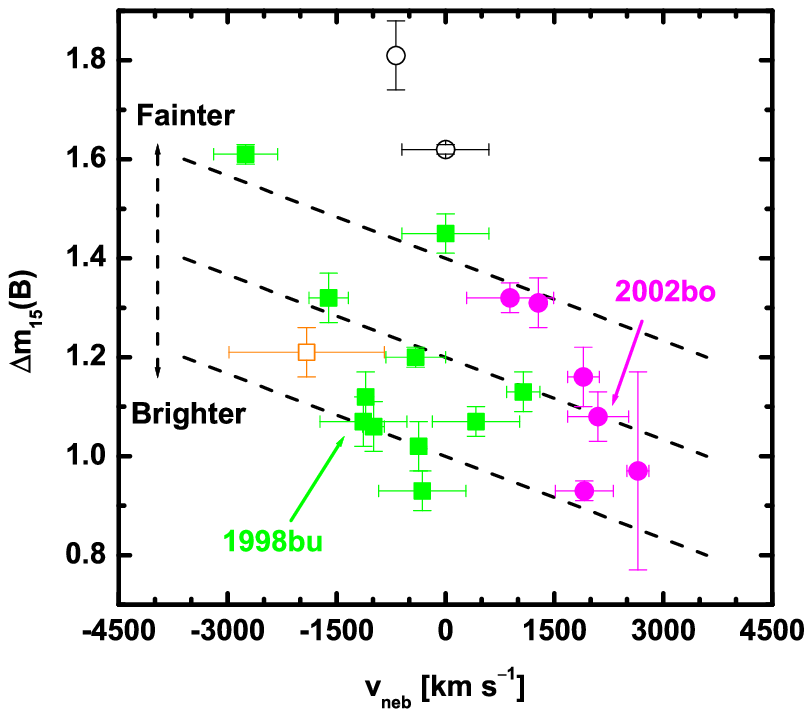}
 \end{center}

The decline-rate parameter $\Delta m_{15} (B)$ versus
the late-time emission-line velocity shift. The three dashed lines 
schematically illustrate the expected viewing angle effect on the 
decline-rate parameter, which would produce a $\sim 0.4$~mag difference 
in the observed $\Delta m_{15} (B)$ for an observer in the 
offset direction as opposed to the opposite direction$^{5}$. 
The lines correspond to three hypothesized explosion configurations 
which are mutually different in $M$($^{56}$Ni)
and therefore in the intrinsic luminosity.

\clearpage
\noindent
31. Phillips, M.~M., et al., 
  The Type Ia Supernova 1986G in NGC 5128 - Optical Photometry and Spectra, 
  Pub. Astron. Soc. Pac., {\bf 99}, 592-605 (1987) \\
\noindent
32. Cristiani, S., et al., 
  The SN 1986 G in Centaurus A, 
  Astron. Astrophys., {\bf 259}, 63-70 (1992) \\
\noindent
33. Leibundgut, B., et al., 
  Premaximum Observations of the Type Ia SN 1990N, 
  Astrophys. J., {\bf 371}, L23-L26 (1991) \\
\noindent
34. G\'omez, G., L\'opez, R., S\'anchez, F., 
  The Canaris Type Ia Supernovae Archive (I), 
  Astron. J., {\bf 112}, 2094-2109 (1996) \\
\noindent
35. Patat, F., et al., 
  The Type Ia Supernova 1994D in NGC 4526: The Early Phases, 
  Mon. Not. R. Astron. Soc., {\bf 278}, 111-124 (1996)\\
\noindent
36. Altavilla, G., et al., 
  Cepheid Calibration of Type Ia Supernovae and The Hubble Constant, 
  Mon. Not. R. Astron. Soc., {\bf 349}, 1344-1352 (2004)\\
37. Riess, A.~G.,  et al., 
  The Rise Time of Nearby Type Ia Supernovae, 
  Astron. J., {\bf 118}, 2675-2688 (1999)\\ 
\noindent
38. Branch, D., et al., 
  Optical Spectra of The Type Ia Supernova 1998aq, 
  Astron. J., {\bf 126}, 1489-1498 (2003)\\
\noindent
39. Hernandez, M., et al., 
  An Early-Time Infrared and Optical Study of The Type Ia Supernova 1998bu in M96, 
  Mon. Not. R. Astron. Soc., {\bf 319}, 223-234 (2000)\\
\noindent
40. Li, W., et al., 
  The Unique Type Ia Supernova 2000cx in NGC 524, 
  Pub. Astron. Soc. Pac., {\bf 113}, 1178-1204 (2001)\\ 
\noindent
41. Patat, F., et al., 
  Upper Limit for Circumstellar Gas around The Type Ia SN 2000cx, 
  Astron. Astrophys., {\bf 474}, 931-936 (2007)\\
\noindent
42. Candia, P., et al., 
  Optical and Infrared Photometry of the Unusual Type Ia Supernova 2000cx, 
  Pub. Astron. Soc. Pac., {\bf 115}, 277-294 (2003)\\ 
\noindent
43. Sollerman, J., et al., 
  The Late-Time Light Curve of The Type Ia Supernova 2000cx, 
  Astron. Astrophys., {\bf 428}, 555-568 (2004)\\
\noindent
44. Krisciunas, K., et al., 
  Optical and Infrared Photometry of the Nearby Type Ia Supernova 2001el, 
  Astron. J., {\bf 125}, 166-180 (2003)\\
\noindent
45. Wang, L., et al., 
  Spectropolarimetry of SN 2001el in NGC 1448: Asphericity of a Normal Type Ia Supernova, 
  Astrophys. J., {\bf 591}, 1110-1128 (2003)\\
\noindent
46. Mattila, S., et al., 
  Early and Late Time VLT Spectroscopy of SN 2001el - 
Progenitor Constraints for a Type Ia Supernova, 
  Astron. Astrophys., {\bf 443}, 649-662 (2005)\\ 
\noindent
47. Pignata, G., et al., 
  Optical and Infrared Observations of SN 2002dj: 
Some Possible Common Properties of Fast-Expanding Type Ia Supernovae, 
  Mon. Not. R. Astron. Soc., {\bf 388}, 971-990 (2008)\\
\noindent
48. Pignata, G., et al., 
  Photometric Observations of The Type Ia SN 2002er in UGC 10743, 
  Mon. Not. R. Astron. Soc., {\bf 355}, 178-190 (2004)\\
\noindent
49. Kotak, R., et al.,
  Spectroscopy of The Type Ia Supernova SN 2002er: Days -11 to +215, 
  Astron. Astrophys., {\bf 436}, 1021-1031 (2005)\\
\noindent 
50. Anupama, G. C., Sahu, D. K., Jose, J., 
  Type Ia Supernova SN 2003du: Optical Observations, 
  Astron. Astrophys., {\bf 429}, 667-676 (2005) \\
\noindent
51. Stanishev, V., et al., 
  SN 2003du: 480 days in The Life of a Normal Type Ia Supernova, 
  Astron. Astrophys., {\bf 469}, 645-661 (2007)\\
\noindent
52. Leloudas, G., et al., 
  The Normal Type Ia SN 2003hv out to Very Late Phases, 
  Astron. Astrophys., {\bf 505}, 265-279 (2009)\\
\noindent
53. Pastorello, A., 
  ESC and KAIT Observations of The Transitional Type Ia SN 2004eo, 
  Mon. Not. R. Astron. Soc., {\bf 377}, 1531-1552 (2007)\\
\noindent
54. Hachinger, S., Mazzali, P.A., \& Benetti, S., 
  Exploring the Spectroscopic Diversity of Type Ia Supernovae, 
  Mon. Not. R. Astron. Soc., {\bf 370}, 299-318 (2006)\\
\noindent
55. Pastorello, A., et al.,  
  ESC Observations of SN 2005cf - I. Photometric Evolution of a Normal Type Ia Supernova, 
  Mon. Not. R. Astron. Soc., {\bf 376}, 1301-1316 (2007)\\
\noindent
56. Wang, X., et al., 
  The Golden Standard Type Ia Supernova 2005cf: Observations from 
The Ultraviolet to The Near-Infrared Wavebands, 
  Astrophys. J., {\bf 697}, 380-408 (2009)\\
\noindent
57. Leonard, D.~C., 
  Constraining the Type Ia Supernova Progenitor: The Search for Hydrogen in Nebular Spectra, 
  AIP Conference Proceedings, {\bf 937}, 311-315 (2007)\\
\noindent
58. Garavini, G., et al., 
   ESC Observations of SN 2005cf. II. Optical Spectroscopy and The High-Velocity Features, 
  Astron. Astrophys., {\bf 471}, 527-535 (2007)\\
\noindent
59. Wang, X. et al., 
  Optical and Near-Infrared Observations of the Highly Reddened, Rapidly Expanding 
Type Ia Supernova SN 2006X in M100, 
  Astrophys. J., {\bf 675}, 626-643 (2008)\\
\noindent
60. Schweizer, F., et al., 
  A New Distance to the Antennae Galaxies (NGC 4038/39) Based on The Type Ia Supernova 2007sr, 
  Astron. J., {\bf 136}, 1482-1489 (2008)\\
\noindent 
61. Motohara, K., et al., 
  The Asymmetric Explosion of Type Ia Supernovae as Seen from Near-Infrared Observations, 
  Astrophys. J., {\bf 652}, L101-L104 (2006)\\
\noindent
62. Gerardy, C.L., et al., 
  Signatures of Delayed Detonation, Asymmetry, and Electron Capture in the Mid-Infrared Spectra of Supernovae 2003hv and 2005df, 
  Astrophys. J., {\bf 661}, 995-1012 (2007)\\
\noindent
63. Filippenko, A.V., et al., 
   The peculiar Type Ia SN 1991T - Detonation of a white dwarf?, 
  Astrophys. J., {\bf 384}, L15-L18 (1992)\\
\noindent
64. Phillips, M. M., et al., 
  SN 1991T - Further evidence of the heterogeneous nature of type Ia supernovae, 
  Astron. J., {\bf 103}, 1632-1637 (1992)\\
\noindent
65. Pakmor, R., et al., 
  Sub-Luminous Type Ia Supernovae from The Mergers of Equal-Mass White Dwarfs 
with Mass $\sim 0.9M_{\odot}$, 
  Nature, {\bf 463}, 61-64 (2010)\\
\noindent
66. Tanaka, M., et al., 
  The Outermost Ejecta of Type Ia Supernovae, 
  Astrophys. J., {\bf 677}, 448-460 (2008)\\ 
\noindent
67. Mazzali, P.A., Sauer, D.N., Pastorello, A., Benetti, S., \& Hillebrandt, W., 
  Abundance Stratification in Type Ia Supernovae - II. The Rapidly Declining, Spectroscopically 
Normal SN 2004eo, 
  Mon. Not. R. Astron. Soc., {\bf 386}, 1897-1906 (2008)\\
\noindent
68. Hachinger, S., Mazzali, P.~A., Taubenberger, S., Pakmor, R., \& Hillebrandt, W., 
  Spectral Analysis of The 91bg-like Type Ia SN 2005bl: 
Low Luminosity, Low Velocities, Incomplete Burning, 
  Mon. Not. R. Astron. Soc., {\bf 399}, 1238-1254 (2009)\\
\noindent 
69. Patat, F., Baade, D., H\"oflich, P., Maund, J. ~R., Wang, L., Wheeler, J. ~C., 
  VLT Spectropolarimetry of The Fast Expanding Type Ia SN 2006X, 
  Astron. Astrophys., {\bf 508}, 229-246 (2009)\\
\noindent
70. Leonard, D.C., Li, W., Filippenko, A.V., Foley, R.J., \& Chornock, R., 
  Evidence for Spectropolarimetric Diversity in Type Ia Supernovae, 
  Astrophys. J., {\bf 632}, 450-475 (2005)\\
\noindent
71. Chornock, R., \& Filippenko, A.V., 
  Deviation from Axisymmetry Revealed by Line Polarization in the Normal 
Type Ia Supernova 2004S, 
  Astron. J., {\bf 136}, 2227-2237 (2008)\\
\noindent
72. Kasen, D., \& Plewa, T., 
  Detonating Failed Deflagration Model of Thermonuclear Supernovae. II. Comparison to Observations, 
  Astrophys. J., {\bf 662}, 459-471 (2007)\\

\end{document}